\documentclass[prl,twocolumn,floatfix,superscriptaddress]{revtex4-1}
\usepackage[pdftex,plainpages=false,colorlinks=true,linkcolor=blue, citecolor=blue, urlcolor=blue]{hyperref}
\usepackage{amsfonts}
\usepackage{amsmath}
\usepackage{amssymb}
\usepackage{graphicx}
\usepackage{natbib}

\begin{document}
 \title{An organizing principle for two-dimensional strongly correlated superconductivity}
\author{L. Fratino}
\affiliation{Department of Physics, Royal Holloway, University of London, Egham, Surrey, UK, TW20 0EX}
\author{P. S\'emon}
\affiliation{D\'epartement de physique and Regroupement qu\'eb\'equois sur les mat\'eriaux de pointe, Universit\'e de Sherbrooke, Sherbrooke, Qu\'ebec, Canada J1K 2R1}
\author{G. Sordi}
\affiliation{Department of Physics, Royal Holloway, University of London, Egham, Surrey, UK, TW20 0EX}
\author{A.-M. S. Tremblay}
\affiliation{D\'epartement de physique and Regroupement qu\'eb\'equois sur les mat\'eriaux de pointe, Universit\'e de Sherbrooke, Sherbrooke, Qu\'ebec, Canada J1K 2R1}
\affiliation{Canadian Institute for Advanced Research, Toronto, Ontario, Canada, M5G 1Z8}

\date{\today}

\begin{abstract}
Superconductivity in the cuprates exhibits many unusual features. 
We study the two-dimensional Hubbard model with plaquette dynamical mean-field theory to address these unusual features and relate them to other normal-state phenomena, such as the pseudogap. 
Previous studies with this method found that upon doping the Mott insulator at low temperature a  pseudogap phase appears. The low-temperature transition between that phase and the correlated metal at higher doping is first-order. A series of crossovers emerge along the Widom line extension of that first-order transition in the supercritical region. 
Here we show that the highly asymmetric dome of the dynamical mean-field superconducting transition temperature $T_c^d$, the maximum of the condensation energy as a function of doping, the correlation between maximum $T_c^d$ and normal-state scattering rate, the change from potential-energy driven to kinetic-energy driven pairing mechanisms can all be understood as remnants of the normal state first-order transition and its associated crossovers that also act as an organizing principle for the superconducting state. 
\end{abstract}
 
\pacs{74.72.-h,71.10.Fd, 74.20.Mn, 71.30.+h}
 
\maketitle


In hole-doped cuprate high-temperature superconductors, d-wave superconductivity shows unusual features that cannot be explained by theoretical methods based on weak correlations~\cite{keimerRev,TremblayJulichPavarini:2013}. 
This has motivated the hypothesis that such unusual features emerge from doping a two-dimensional Mott insulator. 
Advances in this regard were enabled by the development of new theoretical methods such as cluster extensions~\cite{kotliarRMP, maier} of dynamical mean-field theory~\cite{rmp}. A collective effort over the last decade has shown that the key aspects of the phenomenology of cuprates are contained in the two-dimensional Hubbard model. Within this theoretical framework, here we show that these key aspects rest with a single organizing principle, namely a normal-state first-order transition between pseudogap and correlated metal beneath the superconducting dome, identified in Ref.~\cite{ssht}. 
Our analysis indicates that this emerging phase transition at finite doping shapes not only the normal-state phase diagram, but strikingly leaves its mark on the complex structure of the superconducting condensate that is born out of this unusual normal state. 

\begin{figure*}[ht!]
\centering{
\includegraphics[width=0.999\linewidth,clip=]{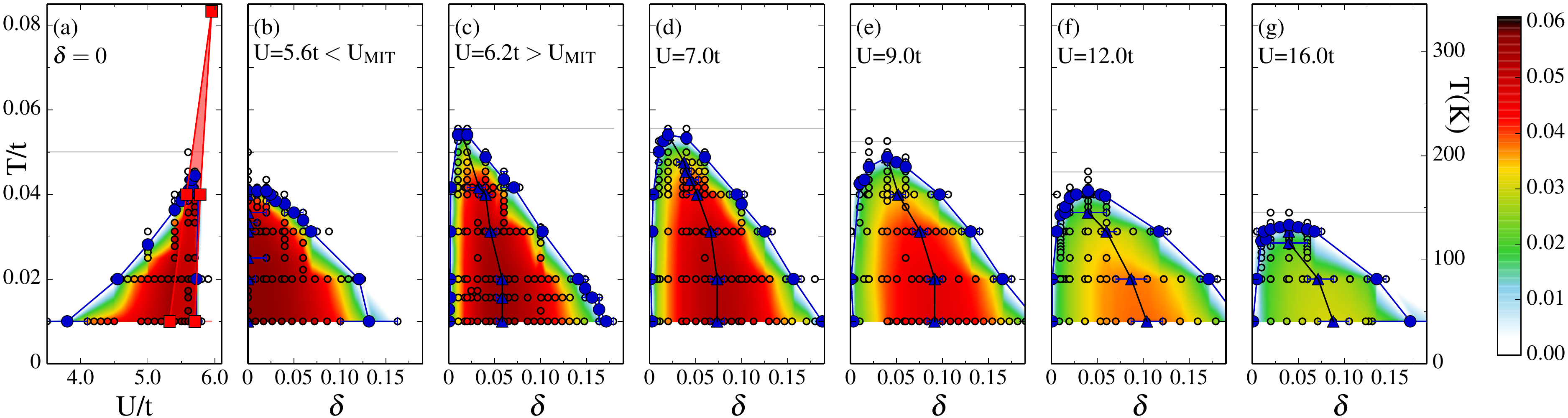}}
\caption{d-wave superconducting phase obtained by the plaquette CDMFT solution of the two-dimensional Hubbard model.  We explore the $T-U-\delta$ space by taking cuts at $n=1$ as a function of $U$ and $T$ [panel (a)] and at constant $U$ as a function of $\delta$ and $T$ [panels (b) to (g)]. Superconductivity is delimited by $T_c^d$ (line with blue filled circles), the temperature below which the superconducting order parameter $\Phi$ is nonzero. Color corresponds to the magnitude of $|\Phi|$ (see supplementary Fig.~S1 for $\Phi(U)$ and $\Phi(\delta)$ curves at different $T$). The loci of $\Phi_{\rm max}(\delta)$ are shown by blue triangles. On the right vertical axis we convert temperature to Kelvin by using $t=0.35$eV. The coexistence region across the first-order Mott metal-insulator transition appears in panel (a) as red shaded area. It is obtained from the hysteretic evolution of the double occupancy with U~\cite{sshtSC}.}
\label{fig1}
\end{figure*}
%


{\it Model and method.}--
The two dimensional Hubbard model on a square lattice reads
\begin{equation}
H=-\sum_{ij\sigma}t_{ij}c_{i\sigma}^\dagger c_{j\sigma}
  +U\sum_{i} n_{i\uparrow } n_{i\downarrow }
  -\mu\sum_{i\sigma} n_{i\sigma}
  \label{eq:HM}
\end{equation}
where $c^{+}_{i\sigma}$ and $c_{i\sigma}$ operators create and destroy an electron of spin $\sigma$ on site $i$, $n_{i\sigma}=c^{+}_{i\sigma}c_{i\sigma}$ is the number operator, $\mu$ is the chemical potential, $U$ the onsite Coulomb repulsion and $t_{ij}$ is the nearest neighbor hopping amplitude.
Neglecting second-neighbor hopping, necessary to capture the correct Fermi surface, minimizes the Monte-Carlo sign-problem and does not alter our main findings (see supplementary Fig.~S7).
Unless specified, the lattice spacing, Planck's constant, Boltzmann's constant and $t$ are unity. 

We solve this model using cellular dynamical mean-field theory~\cite{kotliarRMP, maier} (CDMFT) on a 2$\times$2 plaquette immersed in an infinite self-consistent bath of non-interacting electrons. This plaquette is the minimal cluster that includes all two-dimensional short-range charge, spin and superconducting dynamical correlations. 
We do not take into account long-range charge-density waves in light of the recent experimental results where this transition is removed by pressure~\cite{louis1-2015}. Long-range antiferromagnetism concomitant with long-range superconductivity has been treated at $T$=$0$ in previous work~\cite{davidAF, massimoAF, kancharla}. 
Since we are interested in large values of $U$, i.e. a doped Mott insulator, the most appropriate method to solve the impurity (cluster plus bath) problem is the hybridization expansion continuous-time quantum Monte Carlo method~\cite{millisRMP}. Sign problems prevent the study of large $U$ with alternate quantum Monte Carlo methods~\cite{millisRMP}. 
We use two recent algorithmic improvements to speed up the calculations: a fast rejection algorithm with skip-list data structure~\cite{patrickSkip} and four point updates that are necessary for broken symmetry states like d-wave superconductivity~\cite{patrickERG}.


Let us first consider the superconducting phase diagram. We then discuss features of the normal state that determine its shape. 

{\it Superconducting dome.}--
Previous studies show that both at half-filling and at finite doping the metallic state close to the Mott insulator is unstable to d-wave superconductivity~\cite{maierAF, Lichtenstein:2000,davidAF, kyung:2006, AichhornAFSC:2006, massimoAF, kancharla, Balzer:2010, maierSystem, hauleDOPING,sshtSC,millisSCPG,millisENERGY}. 
In Figure~\ref{fig1} we map out the superconducting state in the $U-T$ plane for the undoped case and in the $\delta-T$ plane for different values of $U$. 
The superconducting region is defined as the region of non-zero superconducting order parameter $\Phi\equiv\langle c_{\mathbf{K}\uparrow}c_{\mathbf{-K}\downarrow} \rangle$ (where the cluster momentum $\mathbf{K}$ is $(\pi,0)$). The boundary, $T_c^d$, is obtained from the mean of the two temperatures where $\Phi$ changes from finite to a small value (here $|\Phi|$=$0.002$). 
While there is no continuous symmetry breaking in two dimensions at finite temperature, $T_c^d$ physically denotes the temperature below which the superconducting pairs form within the cluster~\cite{sshtSC}. 
The actual $T_c$ can be reduced (because of long wavelength thermal or quantum fluctuations~\cite{ekPRL} or of competing long range order~\cite{keimerRev}) or increased (because of pairing through long wavelength antiferromagnetic fluctuations~\cite{Beal-Monod:1986}), but $T_c^d$ still remains a useful quantity marking the region where Mott physics and short-range correlations produce pairing. 

As a function of $U$, $T_c^d$ changes from finite to zero discontinuously at the first-order Mott metal-insulator transition (red shaded region in panel a). Superconductivity appears in the metastable metallic state near the Mott insulator, never in the Mott insulator itself (panels a, b). 
As a function of doping, $T_c^d$ forms a dome as long as $U$ is larger than the critical value necessary to obtain a Mott insulator at half-filling (panels c-g). 
In our previous studies~\cite{sshtSC,patrickERG} we left opened two possibilities: as a function of $\delta$, either superconductivity is separated from the Mott insulator at $\delta=0$ by a first-order transition or there is an abrupt fall of $T_c^d(\delta)$. By increasing the resolution in doping near $\delta=0$, here we find the latter, namely $T_c^d(\delta)$ plummets with decreasing $\delta$. 

The superconducting dome is highly asymmetric. $T_c^d(\delta)$ is zero at $\delta=0$, initially rises steeply with increasing $\delta$, reaching a peak at the optimal doping $\delta_{\rm opt}$ and then declines more gently with further doping. 
The global maximum $T_c^{\rm max}$ of $T_c^d$ in the $U-\delta-T$ space occurs just above $U_{\rm MIT}$ and at finite doping $\delta_{\rm opt}$. Further increase of $U$ leads to a decrease in $T_c^{\rm max}$, as expected if $T_c^d(\delta)$ scales with the superexchange energy $J=4t^2/U$ for large enough $U$~\cite{Kotliar:1988,kancharla}.	 
As a function of $U$, the optimal doping $\delta_{\rm opt}$ departs from $\delta=0$ for $U>U_{\rm MIT}$, increasing with increasing $U$ and saturating around $\delta\approx 0.04$ for large $U$ (see also supplementary Fig.~S2). 

The range of doping where superconductivity occurs at the lowest temperature is consistent~\cite{patrickERG} with results obtained with  CDMFT at $T=0$~\cite{kancharla}. 
The asymmetric superconducting dome with an abrupt fall of $T_c^d$ with decreasing $\delta$ is also consistent with dynamical cluster approximation results on larger clusters~\cite{millisSCPG}. In the latter calculations, the increased accuracy in momentum space leads to a $T_c$ that vanishes before half-filling.

\begin{figure*}[ht!]
\centering{
 \includegraphics[width=0.999\linewidth,clip=]{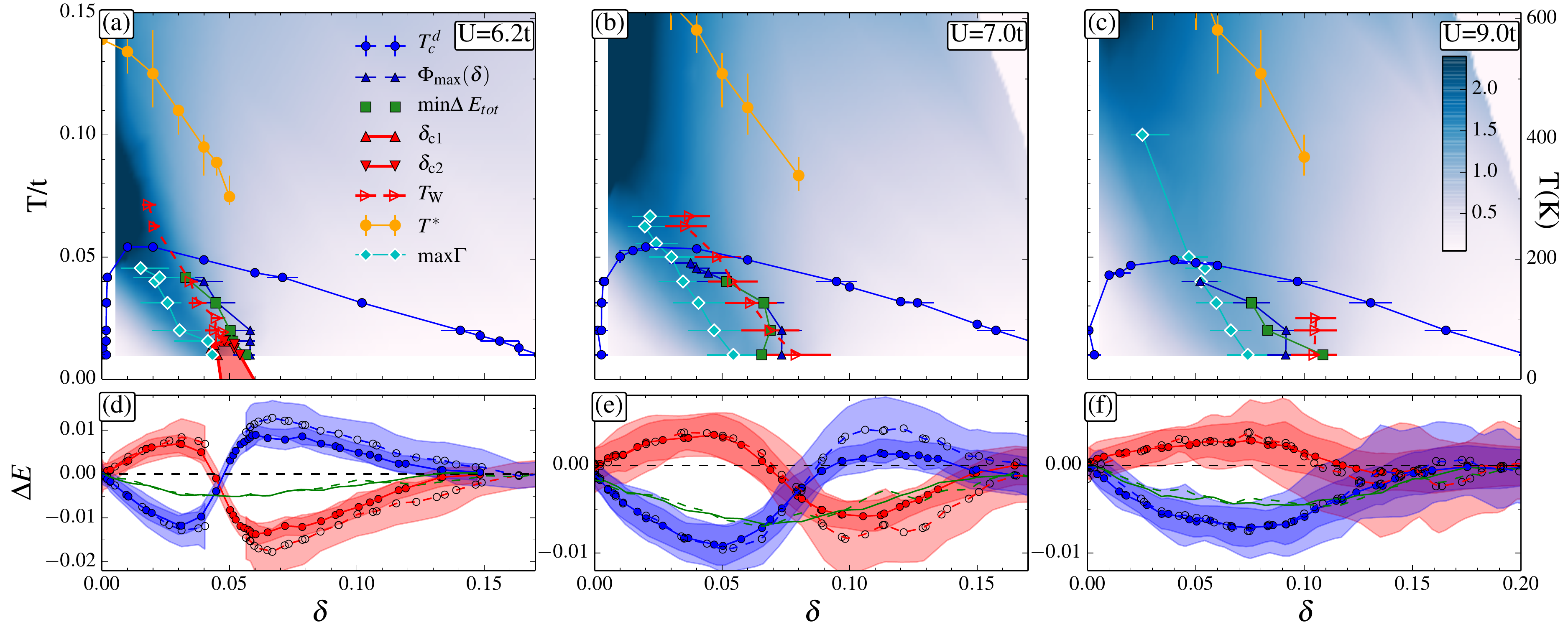}}
\caption{(a), (b), (c): Temperature versus hole doping phase diagram for $U/t=6.2, 7$ and $9$, respectively. Superconductivity is delimited by $T_c^d$ (line with blue filled circles). Beneath the superconducting dome, the normal-state coexistence region across the first-order transition between a pseudogap and a correlated metal appears in (a) as red shaded area. It is delimited by the jumps in the electron density as a function of chemical potential and collapses at the critical endpoint ($T_p,\delta_p$).
The Widom line $T_W$ emerging from the endpoint is estimated by the maxima of the charge compressibility along paths at constant $T$ (line with red triangles)~\cite{ssht}, and the pseudogap onset $T^*$ is computed by the maximum of the spin susceptibility (line with orange circles)~\cite{sshtRHO}. The loci of $\Phi_{\rm max}(\delta)$ are shown by blue triangles and follow $T_W$ of the underlying normal state.  Color corresponds to the magnitude of the scattering rate $\Gamma$, estimated from the zero-frequency extrapolation of the imaginary part of the $(\pi,0)$ component of the cluster self-energy~\cite{sht,sht2}. 
(d), (e), (f): Difference in kinetic, potential and total energies (blue, red and green lines respectively) between the superconducting and normal states, for $T/t=1/50, 1/100$ (full and dashed line, respectively).  Shaded bands give standard errors. The loci where the condensation energy is largest are shown in the upper panels as green filled squares. They follow $T_W(\delta)$ and $\Phi_{\rm max}(\delta)$. }
\label{fig2}
\end{figure*}

{\it Superconducting order parameter.}--
To analyse the shape of the superconducting phase we turn to the superconducting order parameter $\Phi$, whose magnitude is color-coded in Fig.~\ref{fig1} (the raw data is in Fig.~S1).
While  $T_c^{\rm max}$ occurs at finite doping, the overall maximum $\Phi_{\rm max}$ is found in the undoped model close to the Mott insulator. 
But as a function of doping, for $U>U_{\rm MIT}$, $\Phi$ forms a dome that reaches a peak at $\delta_{\Phi_{\rm max}}$.
At our lowest temperature, $\delta_{\Phi_{\rm max}}$ increases with increasing $U$, and saturates around $\delta \approx 0.11$~\cite{kancharla} for large values of $U$. 
Notice that $\delta_{\Phi_{\rm max}}$ at our lowest temperature does not coincide with $\delta_{\rm opt}$, i.e. the doping that optimizes $T_c^d$. Hence, $T_c^d(\delta)$ does not scale with $\Phi(\delta, T\rightarrow 0)$. 
Instead, the locus of the maxima of $\Phi$ in the $\delta-T$ plane at fixed $U$ traces a negatively sloped line within the superconducting dome (lines with blue triangles) that separates the superconducting dome in two regions. 
The sharp asymmetry of the superconducting dome is thus linked to this negatively sloped line, which in turn is related to the phase transition between pseudogap and correlated metal in the underlying normal state, as we discuss below.

{\it Superconductivity and pseudogap.}--
Understanding the normal state has long been considered a prerequisite to a real understanding of high-temperature superconductivity. This comes out clearly from our results.  
Previous normal-state CDMFT studies show that for $U>U_{\rm MIT}$ and small $\delta$, large screened Coulomb repulsion $U$ and the emergent superexchange $J$ lead at low $T$ to a state with strong singlet correlations. That phase has the characteristics of the pseudogap phase~\cite{ssht}. 
The fall of the Knight shift as a function of temperature~\cite{Alloul:1991} is usually associated with $T^*(\delta)$ the onset temperature for the pseudogap. The line with orange filled circles in Figs.~\ref{fig2}a,b,c~\cite{sshtRHO} indicates the onset of the drop of the spin susceptibility and of the density of states as a function of $T$ and the minimum in the $T$ dependence of the c-axis resistivity~\cite{sshtRHO} and is thus $T^*(\delta)$ in our calculation. From our point of view, it is just a precursor to a more fundamental phenomenon. $T^*(\delta)$ exists only if the doping is less than a critical value $\delta<\delta_p$ which is the doping for the critical endpoint ($\delta_p,T_p$) of a first-order transition that appears in Fig.~\ref{fig2}a. A number of crossover lines are associated with this first-order transition. We will discuss them in turn. For larger values of $U$, Fig.~\ref{fig2}b,c, the first-order transition is no-longer visible at accessible temperatures, but the crossovers that are left suggest that it is still present~\cite{sht2}.
 
The normal-state first-order transition separating a pseudogap phase and a correlated metal persists up to the critical endpoint, beyond which only a single normal-state phase exists. Quite generally, different response functions have maxima defining crossover lines emerging from the critical endpoint~\cite{water1}. The Widom line is known as the line where these maxima join asymptotically close to the critical endpoint~\cite{water1}.  
Here we estimate  that line, (red open triangles) $T_{\rm W}$ in the upper panels of Fig.~\ref{fig2}, as the line where the isothermal electronic compressibility has a maximum~\cite{sht,sht2,ssht}. 
Let us briefly consider the other crossover lines. A scan in doping at fixed $T$ shows that the local density of states at the Fermi energy, the spin susceptibility and the c-axis DC conductivity go through an inflection point at $T_{\rm W}(\delta)$~\cite{sshtRHO}. The first-order transition is also a source of anomalous scattering~\cite{sht,sht2}. The blue open diamonds indicate the maximum $\Gamma_{\rm max}$ of the normal state scattering rate $\Gamma$. Its magnitude, estimated from the zero-frequency extrapolation of the imaginary part of the $(\pi,0)$ component of the cluster self-energy, is color-coded in Figs.~\ref{fig2}a,b,c. The region where $\Gamma$ is large is dark blue. It originates at the transition, extends well above $T_c^d$ and is tilted towards the Mott insulator. This large $\Gamma$ is suppressed upon entering the superconducting state~\cite{hauleAVOIDED, hauleDOPING} (see supplementary Fig.~S3). 
 
Even though the first-order transition is absent in the superconducting state, the structure it imposes on the normal state shapes the superconducting phase diagram: 
(a) the maximum of the superconducting order parameter $\Phi_{\rm max}$ (line with blue filled triangles in Figs.~\ref{fig2}a,b,c) parallels $T_W$ and $\Gamma_{\rm max}$, hence the highly asymmetric shape of the superconducting dome is correlated with the slope of the first-order transition and of its supercritical crossovers in the $T-\delta$ plane; 
(b) $\Gamma_{\rm max}$ crosses the superconducting dome approximately at $\delta_{\rm opt}$, hence a region of anomalous scattering broadens as it comes out of the dome; 
(c) since $T^*$ can be detected for doping smaller than $\delta_p $ only, superconductivity and pseudogap are intertwined phenomena: superconductivity can emerge from a pseudogap phase below $\delta_p$, or from a correlated metal above $\delta_p$~\cite{sshtSC}; 
(d) the normal state also controls the source of condensation energy, as we now discuss.

{\it Condensation energy.}--
The superconducting state clearly has a lower free energy than the normal state out of which it is born. In the ground state, the energy difference between both states is known as the condensation energy. The origin of the condensation energy is unambiguous only within a given model~\cite{Chester:1956,LeggettCondEn}. In the BCS model, superconductivity occurs because of a decrease in potential energy. The kinetic energy increase due to particle-hole mixing in the ground state is not large enough to overcome the potential energy drop. In the cuprates, analysis of inelastic neutron scattering~\cite{scalapinoCondEn} has suggested that superconductivity arises because of a gain in exchange energy in the $t-J$ model. Analysis of ARPES~\cite{Norman:2000} and optical data~\cite{molegraaf2002, deutscher2005, carbone2006, giannetti2011} in the context of the Hubbard model has suggested that superconductivity is kinetic-energy driven in the underdoped regime~\cite{andersonBOOK,Hirsch1,scalapinoCondEn, LeggettCondEn, Demler1998}.  

In the lower panels of Fig.~\ref{fig2} we plot, for the Hubbard model Eq.~\ref{eq:HM}, the difference in kinetic and potential energies between the superconducting and normal states ($\Delta E_{\rm kin}$ and $\Delta E_{\rm pot}$; blue and red lines respectively) as a function of doping. The results for the two different temperatures are close enough to suggest we are close to ground state values. The net condensation energy, shown by the green line, is always negative, as expected.    
The doping dependence of $\Delta E_{\rm kin}$ and $\Delta E_{\rm pot}$ on the other hand shows two striking features: it is non monotonic and can display a sign change. For $U=6.2, 7$, Figs.~\ref{fig2}d,e, superconductivity is kinetic-energy driven at small doping and potential energy driven, as in BCS theory, at large doping. 
For $U=9$, Fig.~\ref{fig2}f, superconductivity is kinetic energy driven for all dopings, although the potential energy difference $\Delta E_{\rm pot}$ can change sign.  

Previous investigations~\cite{maierENERGY, carbone2006, millisENERGY} have revealed a complex behavior that remained to this day a puzzle, with $\Delta E_{\rm kin}$ going from negative to positive depending on $T$ and $U$. 
What has been missing to make sense of this complexity is the existence of the normal state first-order transition and its associated supercritical crossovers. By considering different values of $U$, we provide a unified picture of a host of apparently contradictory results. 
For all $U$ considered, the largest condensation energy  (see green line in bottom panels of Fig.~\ref{fig2} and green squares in top panels of Fig.~\ref{fig2}) is concomitant with the largest superconducting order parameter $\Phi_{\rm}(\delta)$ (but not with the maximum $T_c^d$) and hence correlates with the normal-state pseudogap-to-correlated metal first-order transition, and its associated supercritical crossovers. For all $U$, the sign changes are also close to the maximum condensation energy and hence also correlated with the same normal-state features. The influence of Mott and superexchange physics extends unambiguously all the way to the normal-state first-order transition terminating at the critical endpoint, from which supercritical crossovers emerge~\cite{sht}. This reflects itself in the superconducting state in a decisive manner: the changes in sign of the different sources of condensation energy occur for dopings similar to those where the normal-state transition occurs.

{\it Source of condensation energy.}--
Bottom panels of Fig.~\ref{fig2} (see also Fig.~S5) show that in the underdoped region, the kinetic-energy change in the superconducting state is close to minus twice the potential energy change. This is what is expected if superexchange~\cite{FazekasBook:1999} $J$ drives superconductivity there~\cite{Kotliar:1988}. The decrease with $U$  of the maximum $T_c$, of the magnitude of the individual kinetic and potential energy contributions to condensation energy, and of the maximum value of the $T=0$ order parameter~\cite{davidAF,massimoAF,AichhornAFSC:2006,kancharla}, are also all consistent with the importance of $J$ in the effective model that arises from the Hubbard model at large $U$. The BCS-like behavior in the overdoped regime for $U=6.2,7$ probably arises from leftover of the weak-coupling long-wavelength antiferromagnetic spin-wave pairing mechanism~\cite{ScalapinoRev}, although the effect of the self-consistent rearrangement of the spin-fluctuation spectrum in the superconducting state has not been studied yet.

{\it Discussion.}-- 
Our findings further broaden our understanding of the CDMFT solution of the Hubbard model in the doped Mott insulator regime by showing how and to what extent the organizing principle for both the normal state and the superconducting state is the finite-doping first-order transition that determines the shape and the properties of both phases, even though the transition itself is invisible in the superconducting state. In the $T-\delta$ plane, the loci of the maximum order parameter, of the extremum condensation energy, of the maximum normal state scattering relative to the maximum $T_c^d$, all correlate with crossover lines of the underlying normal state that is unstable to d-wave superconductivity.     

We speculate that the application of a magnetic field strong enough to suppress $T_c$ and pressures large enough to remove density waves may reveal the underlying transition. 
We also speculate that sound anomalies associated with the large compressibility in the underlying normal state above the critical endpoint could appear, in analogy with what is observed near the half-filled Mott transition in layered organics~\cite{fournierMott, HassanSound, mck, furukawaWL, vlad, Hebert:2015}. The appearance of large electronic compressibility near the normal state first-order transition suggests that further studies of ubiquitous bond-density waves~\cite{louis1-2015} should be undertaken with the same set of methods.  

We acknowledge D. S\'enechal, L. Taillefer, C. Bourbonnais and H. Alloul for useful discussions. This work was partially supported by the Natural Sciences and Engineering Research council (Canada), and by the Tier I Canada Research Chair Program (A.-M.S.T.). Simulations were performed on computers provided by CFI, MELS, Calcul Qu\'ebec and Compute Canada.

{\bf Author contributions}: L.F. obtained and analysed the data. P.S. wrote the main codes. G.S. and A.-M.S.T. supervised the project and wrote the manuscript, and all authors discussed the results and commented on the manuscript. 

{\bf Competing financial interests}: The authors declare no competing financial interests.


%


\onecolumngrid
\clearpage 
\setcounter{figure}{0}
\makeatletter 
\renewcommand{\thefigure}{S\@arabic\c@figure} 

\begin{center}

{\bf Supplementary information} 

{\bf An organizing principle for two-dimensional strongly correlated superconductivity}

\vspace{0.5cm}

{L. Fratino, P. S\'emon, G. Sordi, A.-M.S. Tremblay}

\end{center}

\vspace{2cm}

In this supplementary information, we first remove ambiguities that might arise from color coding in Fig.~1 and Fig.~2 of the main text by plotting the corresponding raw data, first for the superconducting order parameter in Sec.~A. The location of the maximum $T_c$, of the maximum order parameter and of the end of the superconducting dome as a function of doping is also given. 
The scattering rate $\Gamma$ is in Sec.~B. We also show in this section that the scattering rate decreases drastically in the superconducting state, consistent with the reappearance of quasiparticles in that state. 
Sec.~C summarises the main crossover lines in the normal state found in previous work~\cite{ssht, sshtRHO}. 
We show in Sec.~D how the contribution to the kinetic energy from the plaquette can be isolated from more long-distance related contributions. The plaquette contribution can be computed purely from the $4$ site density matrix. It will be shown that the latter contribution to the condensation energy is always negative, namely the superconducting state always lowers the plaquette kinetic energy.   
Finally, Sec.~E reports the the $T-\delta$ phase diagram for the second neighbor hopping $t'$, to show that main findings of the main text are not altered by $t'$.

\section{A. Superconducting order parameter}
\label{sec:A}

\begin{figure}[!h]
\centering{
\includegraphics[width=1.0\linewidth]{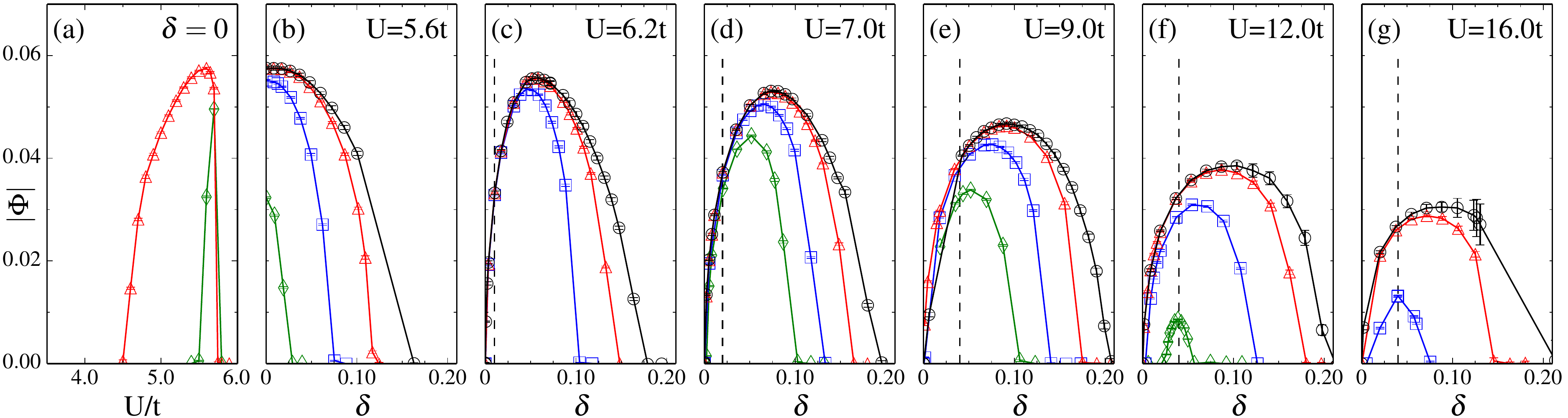}}
\caption{Superconducting order parameter $|\Phi|$ as a function of $U/t$ [panel (a)] and as a function of $\delta$ for several values of the interaction strength $U/t$ [panels (b) to (g)]. The data are shown for temperatures $T/t=1/25$ (green diamonds), $1/32$ (blue squares), $1/50$ (red triangles) and $1/100$ (black circles). Interpolation of these data gives rise to the color map in Fig.~1 of main text. Dashed vertical line displays the optimal doping $\delta_{\rm opt}$. }
\label{figS1}
\end{figure}

\clearpage
\begin{figure}[!ht]
\centering{
\includegraphics[width=0.65\linewidth,clip=]{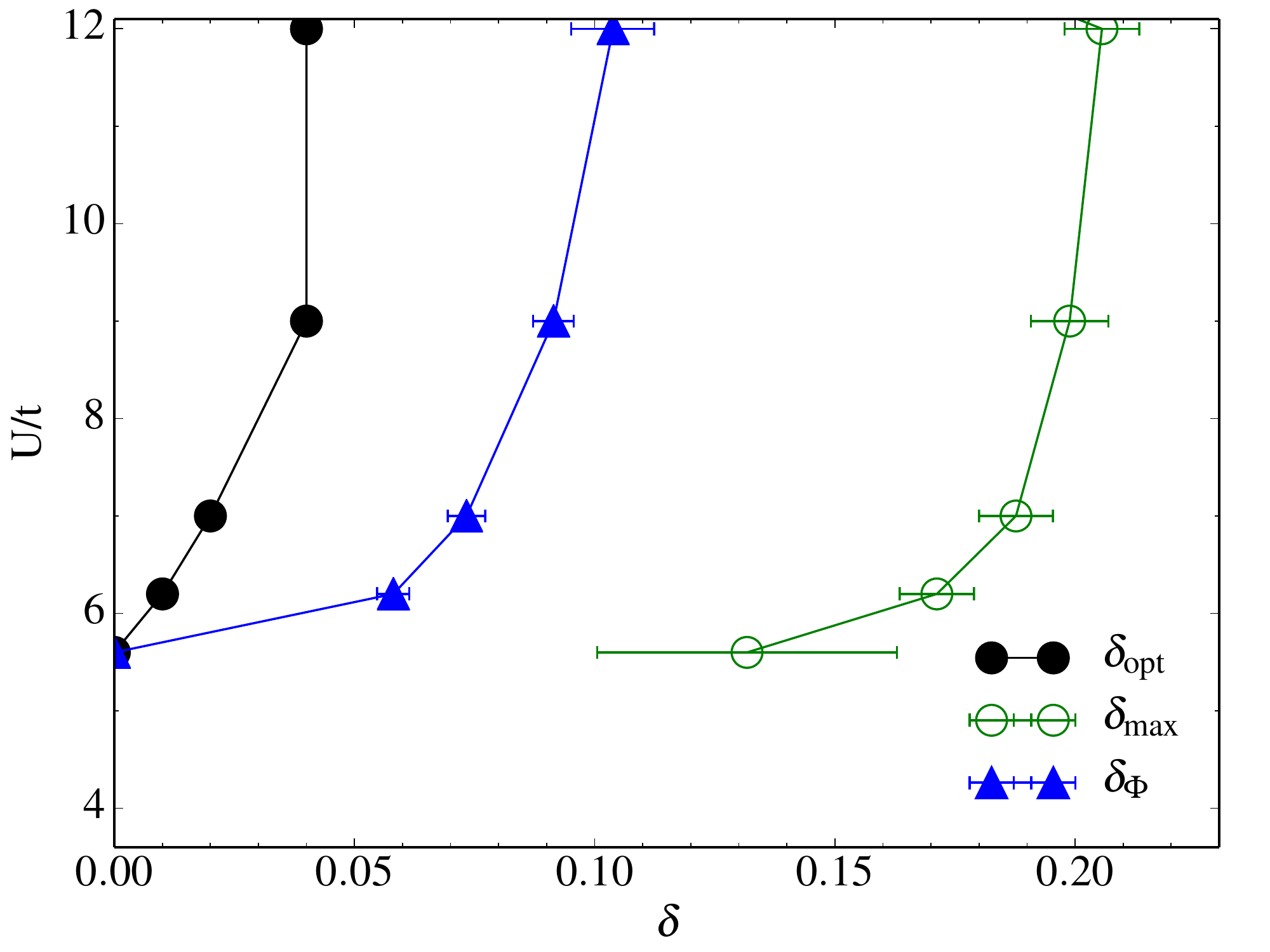}}
\caption{Characteristic dopings in the $U-T$ plane: optimal doping ($\delta_{\rm opt}$, black circles), the position of the maximum order parameter for $T/t=1/100$ ($\delta_{\Phi_{\rm max}}$,  blue triangles) and the largest doping at which superconductivity disappears for the lowest temperature studied, i.e. $T/t=1/100$ ( $\delta_{\rm max}$, green circles). }
\label{figS2}
\end{figure}

\section{B. Scattering rate}
\label{sec:Gamma}

\begin{figure}[!h]
\centering{
\includegraphics[width=1.0\linewidth,clip=]{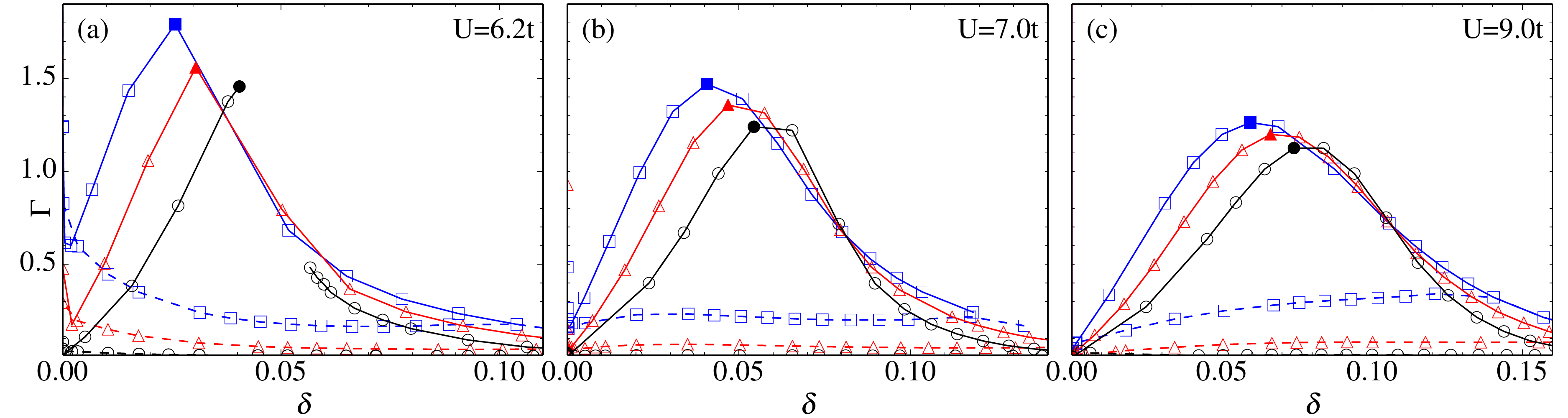}}
\caption{Scattering rate $\Gamma =-{\rm Im}\Sigma_{(\pi,0)}(\omega \rightarrow 0)$ for $U/t=6.2, 7, 9$ in the normal and superconducting states (full and dashed lines, respectively). The data are shown for temperatures $T/t=1/32$ (blue squares), $1/50$ (red triangles) and $1/100$ (black circles). Interpolation of these data gives rise to the color map in top panels of Fig.~2 of the main text. The maximum of the normal state scattering rate $\Gamma(\delta)|_T$ is marked by a solid symbol and is displayed by solid white diamonds in top panels of Fig.~2 of the main text. 
Leaving apart the Mott insulator at $\delta=0$, there is a maximum in the normal state $\Gamma(\delta)|_T$ either close to the first-order transition between pseudogap and correlated metal for $T<T_p$ (cf. $U/t=6.2$ and $T/t=1/100$) or  in the supercritical region for $T>T_p$~\cite{sht,sht2}. Upon increasing temperature, the value of $\Gamma(\delta)|_T$ at its maximum increases as does its width in doping. The large scattering rate is sharply depleted upon entering the superconducting state, as already noticed in Refs.~\cite{hauleDOPING,hauleAVOIDED}.
}
\label{figS3}
\end{figure}

\clearpage
\section{C. Pseudogap to correlated metal transition in the normal state}
\begin{figure}[!h]
\centering{
\includegraphics[width=0.9\linewidth,clip=]{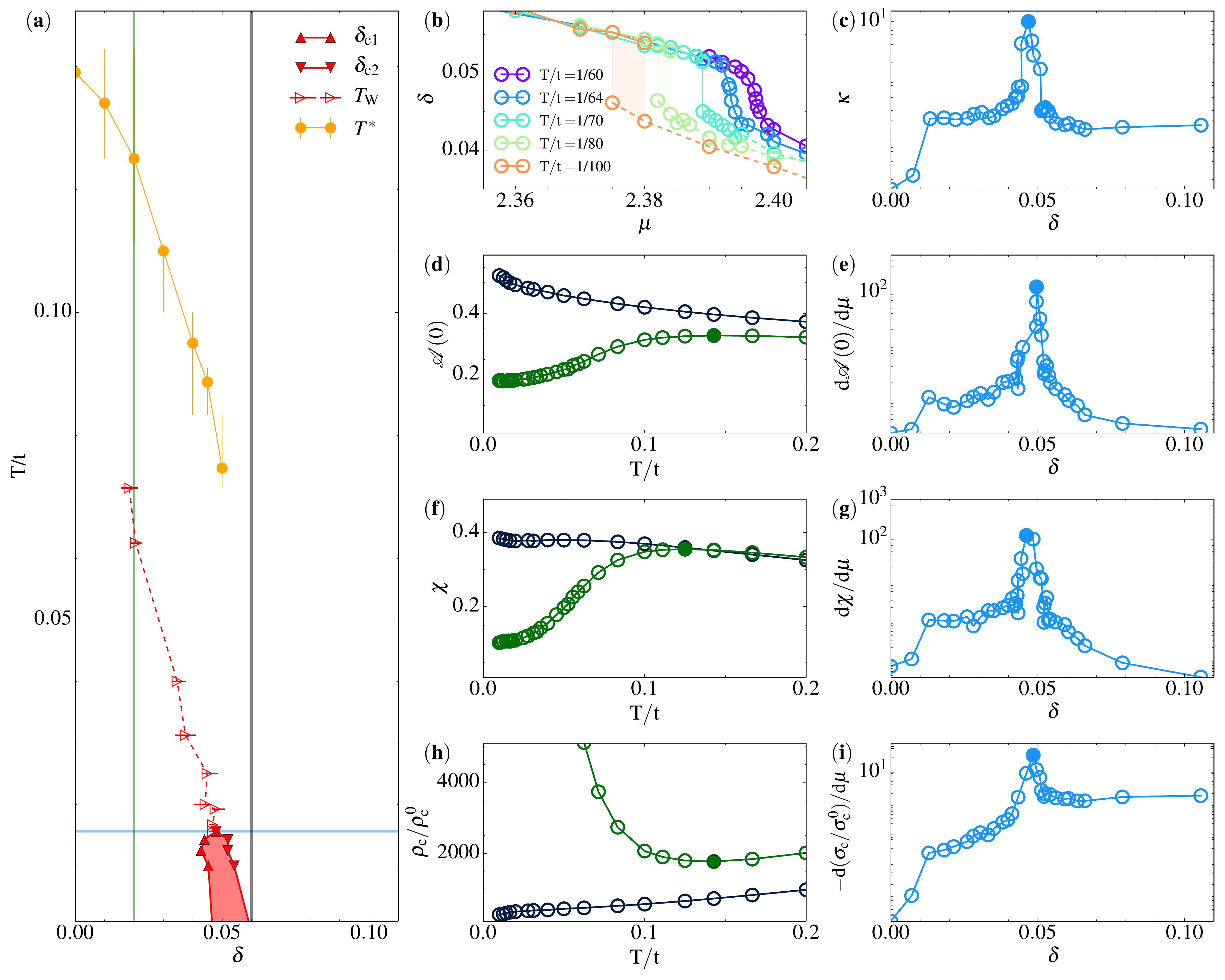}}
\caption{(a) Temperature versus hole doping phase diagram for $U/t=6.2$ in the normal state obtained by CDMFT. Data are taken from our previous investigations~\cite{sht,sht2,ssht,sshtRHO}.  Horizontal (vertical) shaded lines indicate the values of temperature (doping) of the observables in the other panels. 
At zero doping, the system is a Mott insulator and is characterised by a plateau in the occupation at $n(\mu)=1$. At finite doping $\delta=1-n$, the coexistence region across a first-order transition between a pseudogap phase and a correlated metal is shown as red shaded area. Its boundaries are obtained by the jumps in the occupation $n$ versus chemical potential $\mu$ at constant values of temperature, as shown in panel~(b) and discussed in Refs.~\cite{sht,sht2}. Extrapolations to $T=0$ are a guide for the eye. 
The pseudogap to correlated metal first-order transition terminates at a critical endpoint $(\delta_p,T_p) \approx (0.045, 1/65)$. 
Let us first consider paths at constant $T$ [panels (b,c,e,g,i)]. In the supercritical region, $T>T_p$, only one normal-state phase exists and the $n(\mu)$ curves are continuous. The endpoint generates the Widom line $T_W$ (line with red triangles in panel (a)). We estimate $T_W$ by the maxima of the charge compressibility $\kappa=1/n^2(dn/d\mu)_T$, $\rm{max}|_\mu\kappa$~\cite{ssht}. A semilogarithmic plot of $\kappa$ versus $\delta$ at $T/t=1/60$ is shown in panel~(c), and a filled symbol indicates the position of compressibility maximum. The value of $\kappa$ at the maximum increases for $T\rightarrow T_p$, indicating a divergence of $\kappa$ at $T_p$, as investigated in Ref.~\cite{ssht}. The Widom line governs the crossovers of other observables: the local density of states at the Fermi level $\mathcal{A}(\omega=0)$~\cite{ssht}, the spin susceptibility $\chi$~\cite{ssht}, the c-axis DC conductivity $\sigma_c$~\cite{sshtRHO}, all show inflection points as a function of $\mu$. Their derivative with respect to $\mu$ are shown in panels~(e,g,i), respectively. 
Let us now consider scans at constant doping [panels (d,f,h)]. Solely for $\delta<\delta_p$, the temperature dependence of $\mathcal{A}(\omega=0)$, $\chi$ and the c-axis resistivity $\rho_c=1/\sigma_c$ all show non-monotonic behavior. The position of the minima or maxima in such observables is our estimate for the pseudogap onset. For definiteness, we define $T^*$ (line with orange circles in panel (a)) by the maxima in $\chi(T)$.
}
\label{figS6}
\end{figure}

\clearpage

\section{D. Kinetic energy in CDMFT within hybridization expansion impurity solver}\label{sec:kin}

In the hybridization expansion impurity solver, the partition function of the impurity solver is expanded in the hybridization between the impurity and the bath.  
In single-site DMFT~\cite{rmp}, the impurity consists of a site. The kinetic energy per site can be shown~\cite{hauleCTQMC} to be related with the average expansion order by $E_{\rm kin} = -\langle k \rangle /\beta$, where $\beta$ is the inverse temperature. Here we generalize this formula for the CDMFT case. We demonstrate that the kinetic energy is the sum of two terms: similarly to the single-site DMFT case, there is a contribution related to the average expansion order term, but there is another term coming from the cluster (plaquette) part. The latter can be computed from the plaquette density matrix (or occupation numbers). 

The kinetic energy per site reads
\begin{equation}
E_{kin}=\frac{2}{N}\sum_{i,j}\sum_{r,r^{\prime}}t_{ij}\left(  r-r^{\prime
}\right)  \left\langle c_{i}^{\dagger}\left(  r\right)  c_{j}\left(
r^{\prime}\right)  \right\rangle
\end{equation}
where $i,j$ are indices indicating the position within a cluster, $N$ is the number of sites,  and $r, r^{\prime}$ indicate the position of the cluster. 
The sum being on all positions and the hopping matrix $t_{ij}\left(  r-r^\prime\right) $ being symmetric, there is no need to add the hermitian conjugate. 
By inserting the definition of the Green function one obtains 
\begin{equation}
E_{kin}=\frac{T}{N}\sum_{n}e^{-i\omega_{n}0^{-}}\sum_{i,j}\sum_{r,r^{\prime}%
}t_{ij}\left(  r-r^{\prime}\right)  G_{ji}\left(  r^{\prime}-r;i\omega
_{n}\right)  ,
\end{equation}
and by Fourier transformation on the position of the clusters
\begin{equation}
E_{kin}=\frac{2T}{N}\sum_{n}e^{-i\omega_{n}0^{-}}\sum_{i,j}\sum_{\widetilde
{k}}t_{ij}\left(  \widetilde{k}\right)  G_{ji}\left(  \widetilde{k}%
;i\omega_{n}\right)  .
\end{equation}
We keep a discrete wave vector sum. Using the expression for the inverse of the lattice Green function, the hopping can be rewritten so that 
\begin{align}
E_{kin}  &  =\frac{2T}{N}\sum_{n}e^{-i\omega_{n}0^{-}}\sum_{i,j}%
\sum_{\widetilde{k}}\left[  i\omega_{n}+\mu-\Sigma_{ij}\left(  i\omega
_{n}\right)  -G_{ij}\left(  \widetilde{k};i\omega_{n}\right)  ^{-1}\right]
G_{ji}\left(  \widetilde{k};i\omega_{n}\right) \\
&  =\frac{2T}{N}\sum_{n}e^{-i\omega_{n}0^{-}}\left[\sum_{i,j}\sum_{\widetilde{k}%
}\left[  (i\omega_{n}+\mu-\Sigma_{ij}\left(  i\omega_{n}\right)
)G_{ji}\left(  \widetilde{k};i\omega_{n}\right)\right]  -\sum_{i}\sum_{\widetilde{k}}1\right].
\end{align}

The self-consistency condition is given by 
\begin{equation}
G_{ji}^{imp}\left(  i\omega_{n}\right)  =\frac{1}{N_{sr}}\sum_{\widetilde{k}%
}G_{ji}\left(  \widetilde{k};i\omega_{n}\right)  .
\end{equation}
where $N_{sr}=N/N_{c}$, and $N_{c}$ is the cluster size (here $N_{c}=4$). This relation allows one to perform the sum over $\widetilde{k}$ and to write $E_{kin}$ as 
\begin{equation}
E_{kin}=\frac{2T}{N_{c}}\sum_{n}e^{-i\omega_{n}0^{-}}\left[\sum_{i,j}\left[
(i\omega_{n}+\mu-\Sigma_{ij}\left(  i\omega_{n}\right)  )G_{ji}^{imp}\left(
i\omega_{n}\right) \right] -\sum_{i}1\right],
\end{equation}
where we used that $\sum_{\widetilde{k}}=N_{sr}=\frac{N}{N_{c}}$.
Inserting the expression for  $G_{ij}^{imp}\left(  i\omega_{n}\right)  ^{-1}$, one obtains
\begin{align}
E_{kin} &  =\frac{2T}{N_{c}}\sum_{n}e^{-i\omega_{n}0^{-}}\left[\sum_{i,j}\left[
(G_{ij}^{imp}\left(  i\omega_{n}\right)  ^{-1}+t_{ij}^{imp}+\Delta_{ij}\left(
i\omega_{n}\right)  )G_{ji}^{imp}\left(  i\omega_{n}\right) \right]  -\sum_{i}1\right] \\
&  =\frac{2T}{N_{c}}\sum_{n}e^{-i\omega_{n}0^{-}}\sum_{i,j}\left[
(t_{ij}^{imp}+\Delta_{ij}\left(  i\omega_{n}\right)  )G_{ji}^{imp}\left(
i\omega_{n}\right)  \right]  \\
&  =\frac{2T}{N_{c}}\sum_{n}e^{-i\omega_{n}0^{-}}\sum_{i,j}\left[  \Delta
_{ij}\left(  i\omega_{n}\right)  G_{ji}^{imp}\left(  i\omega_{n}\right)
\right]  +\frac{2T}{N_{c}}\sum_{n}e^{-i\omega_{n}0^{-}}\sum_{i,j}\left[
t_{ij}^{imp}G_{ji}^{imp}\left(  i\omega_{n}\right)  \right] . \label{DG+tG}%
\end{align}
Using arguments analogous to those in single-site DMFT,~\cite{hauleCTQMC}  the first term is related to the expansion order~\footnote{The relation to the expansion order is most easily seen in the action formalism.} while the second contribution is
\begin{align}
\frac{2T}{N_{c}}\sum_{n}e^{-i\omega_{n}0^{-}}\sum_{i,j}\left[  t_{ij}%
^{imp}G_{ji}^{imp}\left(  i\omega_{n}\right)  \right]   &  =\frac{2T}{N_{c}%
}\sum_{n}e^{-i\omega_{n}0^{-}}\sum_{K}t_{K}^{imp}G_{K}^{imp}\left(
i\omega_{n}\right)  \\
&  =\frac{1}{N_{c}}\sum_{K}t_{K}^{imp}n_{K}^{imp}.
\end{align}
where $n_{K}^{imp}$ is the occupation of the cluster momentum $K$.  Finally, the total kinetic energy is given by 
\begin{equation}
E_{kin}=-\frac{\left\langle k\right\rangle }{N_{c}\beta}+\frac{1}{N_{c}}%
\sum_{K}t_{K}^{imp}n_{K}^{imp}\label{EcinTot}%
\end{equation}
where $\left\langle k\right\rangle $ is the average expansion order. This last equation serves to define 
\begin{equation}\label{E1,E2}
E_{kin}=E_{kin}^{(1)}+E_{kin}^{(2)}.
\end{equation}
\begin{figure}
\centering{
\includegraphics[width=1.0\linewidth,clip=]{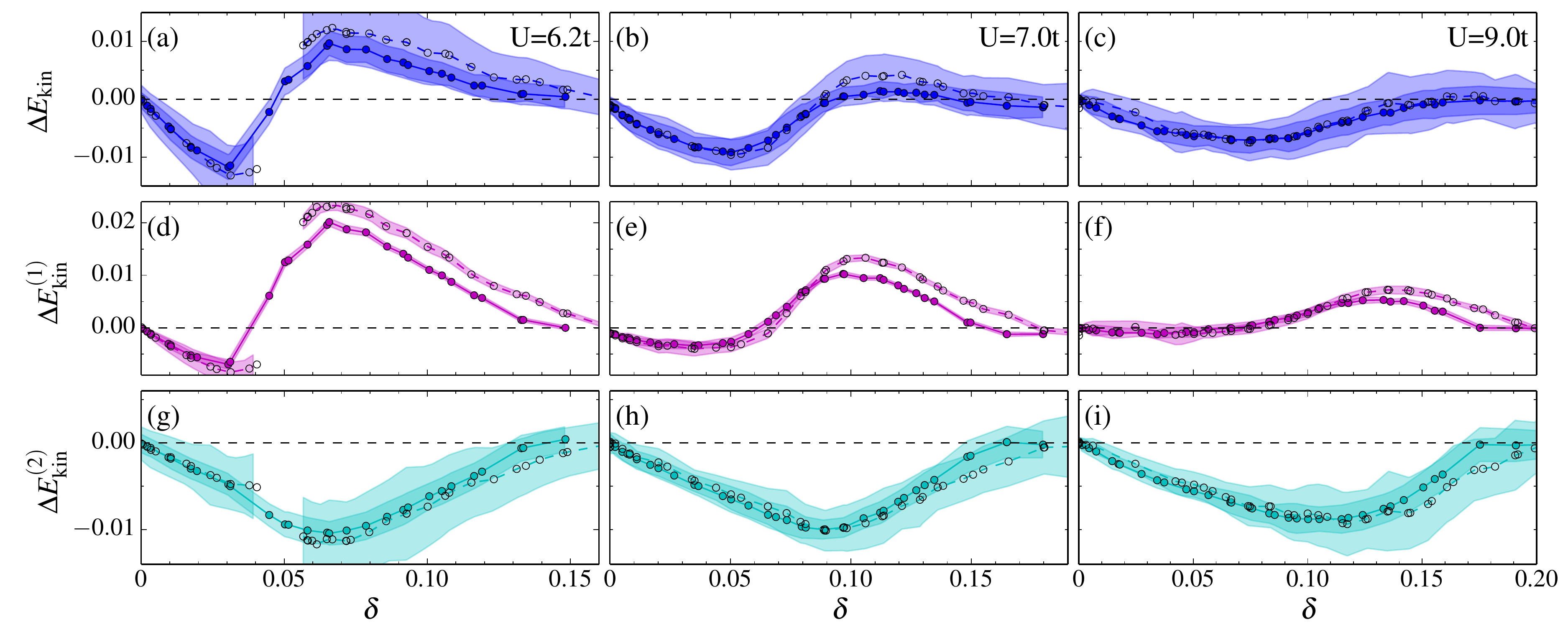}}
\caption{Different contributions to the difference in kinetic energy between the superconducting and the normal state as a function of doping for $U/t=6.2, 7, 9$  (left, central and right columns, respectively) and $T/t = 1/50, 1/100$ (full and dashed line, respectively). Top panels: difference in total kinetic energy $\Delta E_{\rm kin}$; Central panels: contribution from terms outside the cluster  $\Delta E_{\rm kin}^{(1)}$; Bottom panels: contribution from terms within the cluster $\Delta E_{\rm kin}^{(2)}$. We relate the sign change in $\Delta E_{\rm kin}$ to the sign change in $\Delta E_{\rm kin}^{(1)}$. The various contributions are defined by Eqs.~\ref{EcinTot} and \ref{E1,E2}. 
}
\label{figS4}
\end{figure}

The bottom panels in Fig.~\ref{figS4} shows that on short distances, namely within the cluster, the kinetic energy $E_{\rm kin}^{(1)}$ is aways lowered upon entering the superconducting state. However, as the middle panels show, the contribution to the kinetic energy gain coming from longer distance, or smaller wave vectors, can change sign.

\begin{figure}[h!]
	\includegraphics[width=0.95\linewidth,clip=]{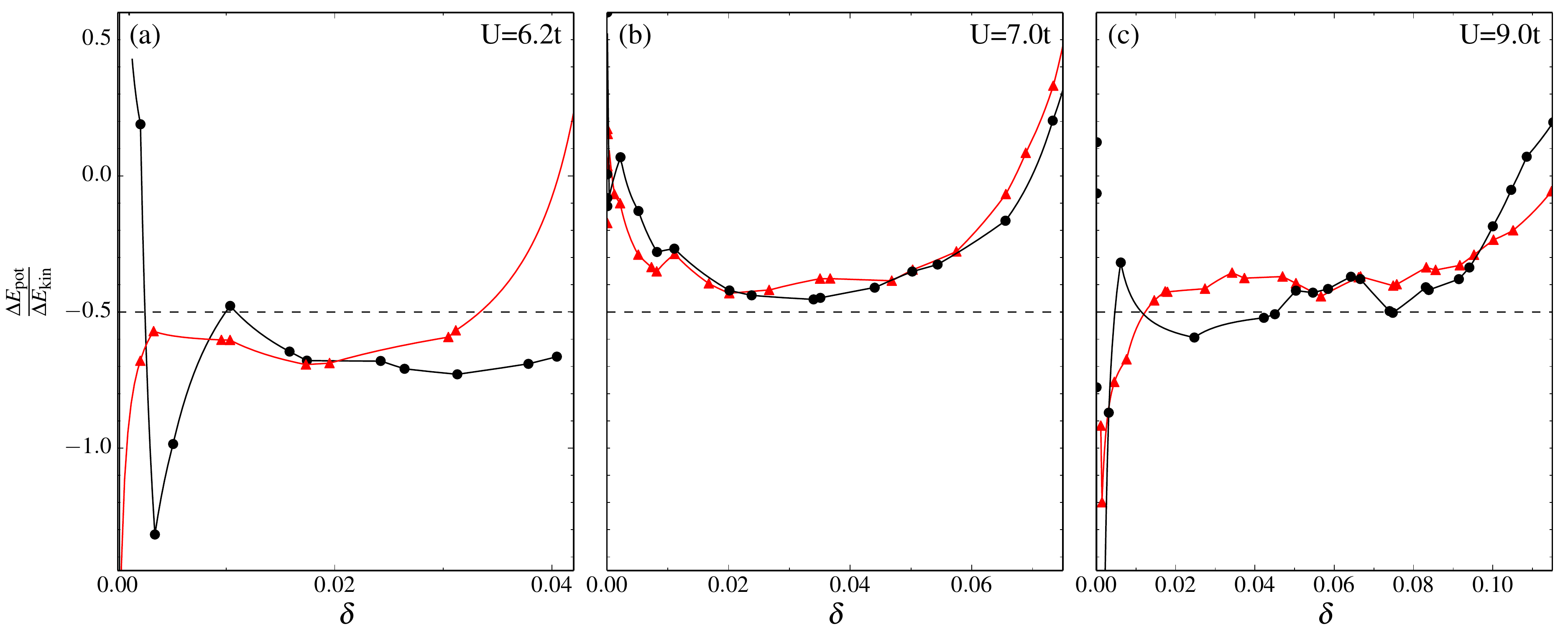}
	\caption{Ratio between potential energy gain and the kinetic energy gain upon entering the superconducting state in the underdoped region, for $T/t = 1/50$ (red triangles) and  $1/100$ (black circles). The horizontal dashed line shows the value $-1/2$ expected from the exchange energy proportional to $J$.}
	\label{figS5}
\end{figure}

Finally, Fig.~\ref{figS5} shows that the ratio between the potential energy gain and the kinetic energy gain is $-1/2$ in the underdoped region. That ratio corresponds to the ratio between potential and kinetic energy contained in the exchange energy, namely the term that scales like $J=4t^2/U$ in the large $U$ limit~\cite{FazekasBook:1999}. It seems that not much energy gain comes from the term in the $t-J$ model describing the hopping of holes. The divergences come from the zero crossings of either the kinetic or the potential energy differences.

\clearpage
%

%

\section{E. Effect of second-neighbor hopping $t'$}
\begin{figure}[h!]
	\centering{
		\includegraphics[width=0.66\linewidth,clip=]{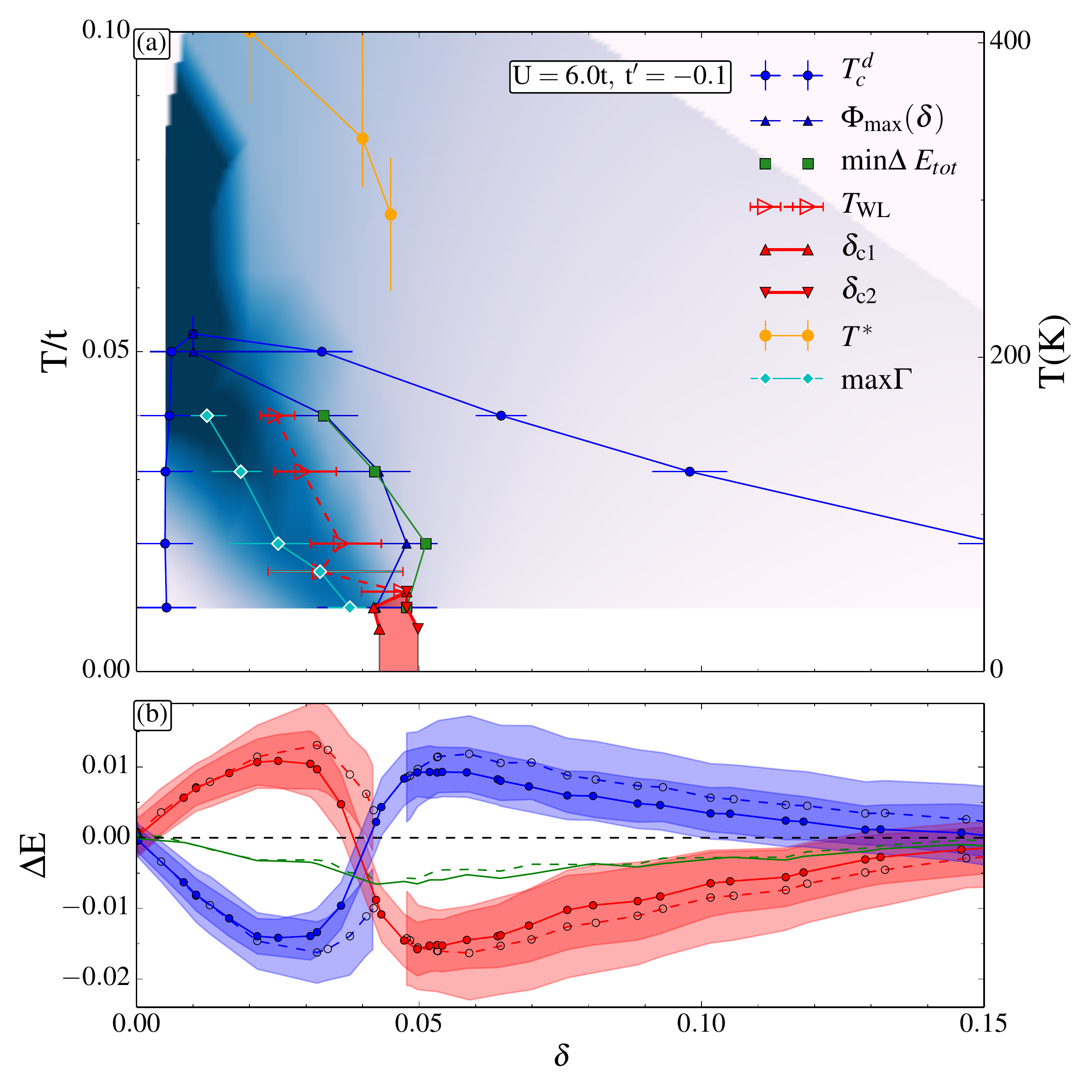}}
	\caption{Same as Fig.~2 of the main text, but for $U=6.0t$ and $t'=-0.10$. All conclusions remain unchanged with a finite $t'$. 
	}
	\label{figS7}
\end{figure}
For $U=0$, the effect of next-nearest-neighbor hopping $t'$ is to move the van Hove singularity to finite doping. This does have some quantitative effect on the phase diagram at finite $U$. However for very large $U$ we expect that this is less important. Given that the sign problem is less severe at $t'=0$ and that values of $U$ can be quite large, the results in the main text are all for $t'=0$. Nevertheless, we performed calculations for $t^\prime=-0.1$, $U=6.0$, which is larger than the critical threshold to open a Mott gap at $n=1$. The results are in Fig.~\ref{figS7}. The value of doping where the first-order transition occurs moves to larger doping, as suggested by Fig. 18 of Ref.~\cite{liebsch}. But one can verify that our qualitative conclusions concerning the organizing principle of the phase diagram are unchanged. The first order transition in the normal state along with the associated crossovers leave their mark in the superconducting state, even though there is no longer a first-order transition in the superconducting state. 

\end{document}